\documentclass[preprintnumbers,superscriptaddress,prb]{revtex4}
\usepackage{amsmath}
\usepackage{amsfonts}
\usepackage{graphicx}
\usepackage{amssymb}

\input{tcilatex}

\begin{document}

\title{Restricted and unrestricted Hartree-Fock approaches for addition
spectrum and Hund's rule of spherical quantum dots in a magnetic field}
\author{C. F. Destefani}
\affiliation{Departamento de F\'{\i}sica, Universidade Federal de S\~{a}o Carlos,
13565-905 S\~{a}o Carlos-SP, Brazil}
\author{J. D. M. Vianna}
\affiliation{Instituto de F\'{\i}sica, Universidade de Bras\'{\i}lia, 70910-900 Bras\'{\i}%
lia-DF, Brazil}
\affiliation{Instituto de F\'{\i}sica, Universidade Federal da Bahia, 40210-340
Salvador-BA, Brazil}
\author{G. E. Marques}
\affiliation{Departamento de F\'{\i}sica, Universidade Federal de S\~{a}o Carlos,
13565-905 S\~{a}o Carlos-SP, Brazil}
\date{\today }

\begin{abstract}
The Roothaan and Pople-Nesbet approaches for real atoms are adapted to
quantum dots in the presence of a magnetic field. Single-particle Gaussian
basis sets are constructed, for each dot radius, under the condition of
maximum overlap with the exact functions. The chemical potential, the
charging energy and the total spin expected values have been calculated, and
we have verified the validity of the quantum dot energy shell structure as
well as the Hund rule for electronic occupation at zero magnetic field. For
finite field, we have observed the violation of Hund's rule and studied the
influence of magnetic field on the closed and open energy shell
configurations. We have also compared the present results with those
obtained with $\mathbf{LS}$-coupling scheme for low electronic occupation
numbers. We focus only on ground state properties and consider quantum dots
populated up to $40$ electrons, constructed by GaAs or InSb semiconductors.
\end{abstract}

\maketitle

\section{Introduction}

The influence of the spatial confinement on the physical properties, such as
electronic spectra of nanostructured systems, is a topic of growing
interest. Among the several kinds of confined systems one can detach low
dimensional electronic gases and impurity atoms in metallic or semiconductor
mesoscopic structures,\cite{jask} as also atoms, ions and molecules trapped
in microscopic cavities,\cite{jask,conne,riv,waz,beekman} where the effects
of the confinement become important when the typical quantum sizes - such as
Fermi wavelength - reach the same order of magnitude as the sizes of the
cavities. However, the energy spectrum of such systems is not only
determined by the spatial confinement and the geometric shape, but also by
environmental factors such as electric and magnetic fields. It is also
defined by many-body effects as electron-electron interaction, which may
even become more important than the confinement itself. In all cases, a
correct description of the physical properties of the problem requires the
system wavefunction to reflect, in an appropriated way, the presence of both
confinement, internal and external interactions, and the corresponding
boundary conditions. We should also mention the existence of other confined
systems, for example, the ones constituted by phonons,\cite{fonon} plasmons%
\cite{plasmon} or confined bosonic gases.\cite{boson}

Low dimensional electronic gases are defined in semiconductor structures
when the bulk translation symmetry is broken in one or more spatial
dimensions, giving origin to 2D (quantum wells), 1D (quantum wires) or 0D
(quantum dots) systems. In such structures, the charged carriers loose, for
some interval of energy, the characteristic of being delocalized in both
three spatial dimensions and become confined in regions of mesoscopic sizes
inside the crystal. This fact transforms the continua energy bands into
broken subbands and energy-gaps, or even into fully discrete energy states
as for example in semiconductor quantum dots (QDs), which are the main type
of confined system to be addressed in this work.

An important point to have in mind is that the usual charging model,\cite%
{11,12,13,14} which reduces the electron-electron interaction inside the dot
to a constant proportional to its electronic occupation, is able to
reasonably reproduce the experimental findings for metallic dots. However,
in order to obtain a more realistic description of many-particle
semiconductor dots, their relatively lower electronic density makes
imperative to consider the electron-electron interaction under a microscopic
point of view. In strong confinement regimes, it may even be included in a
perturbative scheme.

Various are the approaches that have been used to deal with many-particle
QDs. Among them, one can cite charging model, correlated electron model,\cite%
{29} Green functions,\cite{30} Lanczos algorithm,\cite{16} Monte Carlo
method,\cite{31} Hartree-Fock calculations\cite{32,33,34,35} and density
functional theory.\cite{36} It is useful to emphasize that not only the
many-body effects but also the spatial symmetry (geometric confinement)
become indispensable ingredients for a precise determination of quantum
effects in the electronic structure of semiconductor QDs.

Among the several kinds of geometries confining a QD, maybe the most common
be a two-dimensional one defined by a parabolic potential.\cite%
{costa,51,52,53,54} Here we will consider a three-dimensional QD defined by
an infinite spherical potential. The former describes QDs lithographically
defined in the plane of a two-dimensional electron gas, while the latter
describes QDs grown inside glass matrices. Some of the commonly studied
topics in 3D QDs are the formation of energy shells in their spectra,\cite%
{43} the control of electronic correlations,\cite{44} the formation of
Wigner molecules in the system,\cite{45} and the influence of the Coulomb
interaction in their spectra.\cite{46,47} In such spherically defined 3D QD,
both spin and orbital angular momenta are good quantum numbers, and the
many-particle eigenstates can be labelled according to the usual $\mathbf{LS}
$-coupling scheme,\cite{condon} where the exact analytic many-particle
eigenstates are given as a sum over appropriated Slater determinants.
However, in order to deal with QDs having higher occupation, the $\mathbf{LS}
$-coupling is no longer appropriated. Therefore, we have chosen to use in
this paper the Roothaan and Pople-Nesbet matrix formulations\cite{szabo} of
the single determinant self-consistent Hartree-Fock formalism, respectively
appropriated for handling closed and open shell configurations. With them,
we also calculate the QD addition spectrum and show how a magnetic field is
able to violate the Hund rule. Here we consider a QD populated with up to $%
N=40$ electrons under the presence of a magnetic field, and expand our basis
in a set of properly optimized Gaussian functions. As a note one could, in
principle, also include spin-orbit coupling in the model, where $L$ and $S$
would then no longer be good quantum numbers, but for an infinite spherical
potential such coupling yields no contribution to the total energy of the
system.

The paper is organized as follows. In section II, for completeness and in
order to introduce our notation, both restricted and unrestricted formalisms
are resumed. In section III we show the Hamiltonian model and discuss how
both QD chemical potential and charging energy are calculated; also, we
present details on how the inclusion of a magnetic field changes the
previous formalisms, as well as how the Gaussian basis set is constructed.
Finally, in section IV we show our results and we use section V to line up
our conclusions.

\section{Theoretical method}

The Hartree-Fock (HF) approach assumes that the $N$-electron ground state of
an interacting system is given by the single Slater determinant $\left\vert
\Psi _{0}\right\rangle =\left\vert \chi _{1}\chi _{2}...\chi _{a}\chi
_{b}...\chi _{N-1}\chi _{N}\right\rangle $, where the set of optimized spin
orbitals $\left\{ \chi _{a}|a=1...N\right\} $ is obtained by the
minimization of the total energy $E_{0}$ of such state, 
\begin{equation}
E_{0}=\left\langle \Psi _{0}\right\vert H\left\vert \Psi _{0}\right\rangle
=\sum_{a}h_{aa}+\frac{1}{2}\sum_{a}\sum_{b}\left[ J_{ab}-K_{ab}\right] \text{%
,}  \label{single}
\end{equation}%
where the kinetic $h_{aa}$, direct $J_{ab}$, and exchange $K_{ab}$
contributions to $E_{0}$ are given by 
\begin{gather}
h_{aa}=\left\langle \chi _{a}\right\vert h_{a}\left\vert \chi
_{a}\right\rangle =-\frac{\hbar ^{2}}{2m}\int d^{4}\mathbf{x}_{1}\chi
_{a}^{\ast }(1)\left[ \nabla ^{2}\chi _{a}(1)\right] \text{,}  \notag \\
J_{ab}=\left\langle \chi _{a}\right\vert J_{b}\left\vert \chi
_{a}\right\rangle =\frac{e^{2}}{\varepsilon }\int d^{4}\mathbf{x}_{1}\chi
_{a}^{\ast }(1)\left[ \int d^{4}\mathbf{x}_{2}\chi _{b}^{\ast }(2)\frac{1}{%
\left\vert \mathbf{r}_{1}-\mathbf{r}_{2}\right\vert }\chi _{b}(2)\right]
\chi _{a}(1)\text{,}  \label{integrals} \\
K_{ab}=\left\langle \chi _{a}\right\vert K_{b}\left\vert \chi
_{a}\right\rangle =\frac{e^{2}}{\varepsilon }\int d^{4}\mathbf{x}_{1}\chi
_{a}^{\ast }(1)\left[ \int d^{4}\mathbf{x}_{2}\chi _{b}^{\ast }(2)\frac{1}{%
\left\vert \mathbf{r}_{1}-\mathbf{r}_{2}\right\vert }\chi _{a}(2)\right]
\chi _{b}(1)\text{,}  \notag
\end{gather}%
with $m$ ($\varepsilon $) being the effective electron mass (material
dielectric constant), and $\nabla ^{2}=r^{-2}(\partial /\partial
r(r^{2}\partial /\partial r)-L^{2})$ being the Laplacian. The symbol $d^{4}%
\mathbf{x}$ is used in the integrals because the orbitals $\left\{ \chi
_{a}\right\} $ involve both spatial and spin parts. The minimization of Eq. (%
\ref{single}) yields the self-consistent integro-differential HF equation,
whose procedure in order to obtain the optimized spin orbitals can be found
elsewhere;\cite{szabo} it is given by $f\left\vert \chi _{a}\right\rangle
=\varepsilon _{a}\left\vert \chi _{a}\right\rangle $, where the Fock
operator is 
\begin{equation}
f=h_{a}+\sum_{b}\left[ J_{b}-K_{b}\right] \text{.}  \label{fock}
\end{equation}%
It can be straightly verified that the application of $f$ on $\left\vert
\chi _{a}\right\rangle $ can be obtained from Eq. (\ref{integrals}). The
ground state energy $E_{0}$ is calculated in each iteration of our code
until a convergence of $10^{-9}$ eV be reached.

The above expressions are general in the sense that no supposition was taken
on the spin orbitals $\left\{ \chi _{a}\right\} $. Furthermore, the
confinement potential $V(\mathbf{r})$ should be added to $H$, however, since
we have chosen to work with an infinite spherical potential barrier, we have
excluded it from our expressions.

\subsection{Roothaan approach}

The first supposition, which is the core of the Restricted Hartree-Fock
approach (RHF), consists in assuming that both $\alpha $ (up) and $\beta $
(down) spin functions are restricted to have the same spatial function or,
more specifically, $\chi _{i}(\mathbf{x})=\left\{ \psi _{j}(\mathbf{r}%
)\alpha (\omega ),\psi _{j}(\mathbf{r})\beta (\omega )\right\} $. In such
way, the ground state of the system becomes $\left\vert \Psi
_{0}^{RHF}\right\rangle =\left\vert \psi _{1}\overline{\psi }_{1}...\psi _{a}%
\overline{\psi }_{a}...\psi _{N/2}\overline{\psi }_{N/2}\right\rangle $,
where a spatial function with (without) upper bar labels the spin-down
(spin-up) state. Therefore, within the RHF approach, the spatial orbitals $%
\left\{ \psi _{a}|a=1...N/2\right\} $ are doubly occupied and $\left\vert
\Psi _{0}^{RHF}\right\rangle $ represents a closed shell system having total
angular momenta $J=L=S=0$. After integration over the spin degree of freedom,%
\cite{szabo} the HF equation in the restricted formalism becomes $%
f\left\vert \psi _{a}\right\rangle =\varepsilon _{a}\left\vert \psi
_{a}\right\rangle $, where the Fock operator is given by 
\begin{equation}
f=h_{a}+\sum_{b}^{N/2}\left[ 2J_{b}-K_{b}\right] \text{,}  \label{fo_fe}
\end{equation}%
and the ground state energy becomes 
\begin{equation}
E_{0}^{RHF}=2\sum_{a}^{N/2}h_{aa}+\sum_{a}^{N/2}\sum_{b}^{N/2}\left[
2J_{ab}-K_{ab}\right] \text{.}  \label{E0_fe}
\end{equation}%
Here, $h_{aa}=\left\langle \psi _{a}\right\vert h_{a}\left\vert \psi
_{a}\right\rangle $, $J_{ab}=\left\langle \psi _{a}\right\vert
J_{b}\left\vert \psi _{a}\right\rangle $, and $K_{ab}=\left\langle \psi
_{a}\right\vert K_{b}\left\vert \psi _{a}\right\rangle $; so, the
integrations in Eq. (\ref{integrals}) involve only spatial functions in the
RHF approach.

The Roothaan formalism consists in translating the RHF equations into a
matrix formulation. This is achieved by expanding the set $\left\{ \psi
_{i}|i=1...k\right\} $ to be minimized, in a set of known basis functions $%
\left\{ \phi _{\nu }|\nu =1...k\right\} $, 
\begin{equation}
\psi _{i}=\sum_{\nu }C_{\nu i}\phi _{\nu }\text{,}  \label{exp}
\end{equation}%
where the spatial orbitals having $k>N/2$ are empty. Within this procedure,
the coefficients $C_{\nu i}$ become the parameters to be iterated. When Eq. (%
\ref{exp}) is inserted in the closed shell HF equation ($f$ given in Eq. (%
\ref{fo_fe})) and multiplied from the left by $\phi _{\mu }^{\ast }(\mathbf{r%
})$, one obtains the Roothaan characteristic $k$ x $k$ matrix expression, 
\begin{equation}
\mathbf{FC}=\mathbf{SC\varepsilon }\text{,}  \label{bic}
\end{equation}%
where $\mathbf{S}$ is the positive defined overlap matrix between the basis
functions, whose elements are 
\begin{equation}
S_{\mu \nu }=\int d^{3}\mathbf{r}\phi _{\mu }^{\ast }(\mathbf{r})\phi _{\nu
}(\mathbf{r})\text{,}  \label{over}
\end{equation}%
$\mathbf{C}$ is the matrix of the expansion coefficients $C_{\nu i}$, whose
columns describe each spatial orbital $\psi _{i}$, $\mathbf{\varepsilon }$
is the diagonal matrix of the orbital energies $\varepsilon _{i}$, and $%
\mathbf{F}$ is the matrix of the Fock operator $f$, whose elements are 
\begin{equation}
F_{\mu \nu }=\int d^{3}\mathbf{r}\phi _{\mu }^{\ast }(\mathbf{r})f\phi _{\nu
}(\mathbf{r})\text{.}  \label{F}
\end{equation}%
In order to explicitly write the Fock matrix, it is convenient to introduce
the charge density of the system. For a closed shell configuration,
described by a single determinant where each spatial orbital is doubly
occupied, it can be written as $\rho (\mathbf{r})=2\sum_{a}^{N/2}\left\vert
\psi _{a}(\mathbf{r})\right\vert ^{2}=\sum_{\mu }\sum_{\nu }P_{\mu \nu }\phi
_{\mu }(\mathbf{r})\phi _{\nu }^{\ast }(\mathbf{r})$, where one defines the
density matrix $\mathbf{P}$ whose elements have the form 
\begin{equation}
P_{\mu \nu }=2\sum_{a}^{N/2}C_{\mu a}C_{\nu a}^{\ast }\text{;}  \label{pp_fe}
\end{equation}%
observe that an integration of $\rho (\mathbf{r})$ over all space yields $N$%
. The elements of the Fock matrix are obtained when one inserts Eq. (\ref%
{fo_fe}) into Eq. (\ref{F}) and uses Eqs. (\ref{exp}) and (\ref{pp_fe}).
After some algebra, one gets $F_{\mu \nu }=T_{\mu \nu }+G_{\mu \nu }$, where
the kinetic contribution is given by 
\begin{equation}
T_{\mu \nu }=-\frac{\hbar ^{2}}{2m}\int d^{3}\mathbf{r}\phi _{\mu }^{\ast }(%
\mathbf{r})\left[ \nabla ^{2}\phi _{\nu }(\mathbf{r})\right] \text{,}
\label{kin}
\end{equation}%
while the Coulomb contribution is written as 
\begin{eqnarray}
G_{\mu \nu } &=&\frac{e^{2}}{\varepsilon }\sum_{\lambda }\sum_{\sigma
}P_{\lambda \sigma }\left[ \int d^{3}\mathbf{r}_{1}\int d^{3}\mathbf{r}%
_{2}\phi _{\mu }^{\ast }(\mathbf{r}_{1})\phi _{\sigma }^{\ast }(\mathbf{r}%
_{2})\frac{1}{\left\vert \mathbf{r}_{1}-\mathbf{r}_{2}\right\vert }\phi
_{\nu }(\mathbf{r}_{1})\phi _{\lambda }(\mathbf{r}_{2})\right.   \notag \\
&&\left. -\frac{1}{2}\int d^{3}\mathbf{r}_{1}\int d^{3}\mathbf{r}_{2}\phi
_{\mu }^{\ast }(\mathbf{r}_{1})\phi _{\sigma }^{\ast }(\mathbf{r}_{2})\frac{1%
}{\left\vert \mathbf{r}_{1}-\mathbf{r}_{2}\right\vert }\phi _{\lambda }(%
\mathbf{r}_{1})\phi _{\nu }(\mathbf{r}_{2})\right] \text{.}  \label{coulr}
\end{eqnarray}%
The self-consistency of this method appears as the dependence of $\mathbf{F}$
on $\mathbf{P}$ that, in turns, depends on $\mathbf{C}$, which are the
parameters to be determined.

Therefore, the procedure for the solution of Eq. (\ref{bic}) must follow the
steps: i) Given a confinement potential for the system, one specifies $N$
and $\left\{ \phi _{\mu }\right\} $; ii) The integrations in $S_{\mu \nu }$
and $T_{\mu \nu }$ are done; iii) An initial guess is used for $\mathbf{P}$;
iv) With $\mathbf{P}$ and the two-electron integrals one obtains the matrix $%
\mathbf{G}$, which is added to the matrix $\mathbf{T}$ (once calculated,
this one-electron matrix no longer changes, since it does not depend on $%
\mathbf{P}$ or $\mathbf{C}$) to form $\mathbf{F}$; v) $\mathbf{F}$ is
diagonalized in order to get $\mathbf{C}$ and $\mathbf{\varepsilon }$, and
the solution is used to form a new matrix $\mathbf{P}$ from Eq. (\ref{pp_fe}%
); vi) This iteration is performed until the desired convergence be found.
The criterion for convergence may also be set in the change of the ground
state energy that is calculated in each iteration; in the Roothaan formalism
it can be written as 
\begin{equation}
E_{0}^{RHF}=\frac{1}{2}\sum_{\mu }\sum_{\nu }P_{\nu \mu }\left( T_{\mu \nu
}+F_{\mu \nu }\right) \text{.}  \label{E0_root2}
\end{equation}

\subsection{Pople-Nesbet approach}

The second supposition is based on the relaxation of Roothaan's restriction
by letting the $\alpha $ and $\beta $ spin functions have different spatial
components, which is the core of the Unrestricted Hartree-Fock approach
(UHF). Thus, $\chi _{i}(\mathbf{x})=\left\{ \psi _{j}^{\alpha }(\mathbf{r}%
)\alpha (\omega ),\psi _{j}^{\beta }(\mathbf{r})\beta (\omega )\right\} $,
where spin-up (spin-down) electrons are described by the spatial orbitals $%
\left\{ \psi _{j}^{\alpha }|j=1...k\right\} $ ($\{\psi _{j}^{\beta
}|j=1...k\}$). Within the UHF formalism, an unrestricted wavefunction has
the form $\left\vert \Psi ^{UHF}\right\rangle =\left\vert \psi _{1}^{\alpha }%
\overline{\psi }_{1}^{\beta }...\right\rangle $, which represents an open
shell system since no spatial orbital can be doubly occupied. These UHF
functions are not necessarily eigenstates of the system having defined $L$
and $S$ values, and $N$ is no longer restricted to an even number, but it
has to satisfy the condition $N=N^{\alpha }+N^{\beta }$, the sum of spin-up
and spin-down electrons.

In analogy with the RHF case, the spin can be integrated in the UHF
formalism.\cite{szabo} The main difference here is that there are two
coupled HF equations to be simultaneously solved. They are given by $%
f^{\alpha /\beta }\left\vert \psi _{j}^{\alpha /\beta }\right\rangle
=\varepsilon _{j}^{\alpha /\beta }\left\vert \psi _{j}^{\alpha /\beta
}\right\rangle $, where the respective Fock operators are 
\begin{equation}
f^{\alpha /\beta }=h_{j}+\sum_{a}^{N^{\alpha /\beta }}\left[ J_{a}^{\alpha
/\beta }-K_{a}^{\alpha /\beta }\right] +\sum_{a}^{N^{\beta /\alpha
}}J_{a}^{\beta /\alpha }\text{.}  \label{fafb2}
\end{equation}%
Notice that this equation reduces to Eq. (\ref{fo_fe}) if $\psi _{j}^{\alpha
}=\psi _{j}^{\beta }$. The way in which the corresponding terms in $%
f^{\alpha /\beta }$ operate on $\left\vert \psi _{j}^{\alpha /\beta
}\right\rangle $ can be obtained from Eq. (\ref{integrals}), keeping in mind
that here one deals with only spatial contributions. In both $f^{\alpha }$
and $f^{\beta }$ there is presence of a kinetic term $h_{j}$, a direct $%
J_{a}^{\alpha /\beta }$ and an exchange $K_{a}^{\alpha /\beta }$ term
between electrons with same spin, and a direct term $J_{a}^{\beta /\alpha }$
between electrons with different spin. Observe that it is such
interdependence among $f^{\alpha }$ and $\psi _{j}^{\beta }$, as well as
among $f^{\beta }$ and $\psi _{j}^{\alpha }$, that makes necessary a
simultaneous solution of the two HF equations in the UHF approach. Such
solution yields the sets $\left\{ \psi _{j}^{\alpha }\right\} $ and $\left\{
\psi _{j}^{\beta }\right\} $ that minimize the ground state energy $%
E_{0}^{UHF}$ of the unrestricted state $\left\vert \Psi
_{0}^{UHF}\right\rangle $, given by 
\begin{equation}
E_{0}^{UHF}=\sum_{a}^{N^{\alpha }}h_{aa}^{\alpha }+\sum_{a}^{N^{\beta
}}h_{aa}^{\beta }+\frac{1}{2}\sum_{a}^{N^{\alpha }}\sum_{b}^{N^{\alpha }}%
\left[ J_{ab}^{\alpha \alpha }-K_{ab}^{\alpha \alpha }\right] +\frac{1}{2}%
\sum_{a}^{N^{\beta }}\sum_{b}^{N^{\beta }}\left[ J_{ab}^{\beta \beta
}-K_{ab}^{\beta \beta }\right] +\sum_{a}^{N^{\alpha }}\sum_{b}^{N^{\beta
}}J_{ab}^{\alpha \beta }\text{.}  \label{eo1}
\end{equation}%
Here, $h_{aa}^{\alpha /\beta }=\left\langle \psi _{a}^{\alpha /\beta
}\right\vert h_{a}\left\vert \psi _{a}^{\alpha /\beta }\right\rangle $, $%
J_{ab}^{\alpha \beta }=\left\langle \psi _{a}^{\alpha }\right\vert
J_{b}^{\beta }\left\vert \psi _{a}^{\alpha }\right\rangle =\left\langle \psi
_{b}^{\beta }\right\vert J_{a}^{\alpha }\left\vert \psi _{b}^{\beta
}\right\rangle =J_{ba}^{\beta \alpha }$, $J_{ab}^{\alpha \alpha
}=\left\langle \psi _{a}^{\alpha }\right\vert J_{b}^{\alpha }\left\vert \psi
_{a}^{\alpha }\right\rangle =\left\langle \psi _{b}^{\alpha }\right\vert
J_{a}^{\alpha }\left\vert \psi _{b}^{\alpha }\right\rangle =J_{ba}^{\alpha
\alpha }$, plus the analog term for $J_{ab}^{\beta \beta }$. Also, $%
K_{ab}^{\alpha \alpha }=\left\langle \psi _{a}^{\alpha }\right\vert
K_{b}^{\alpha }\left\vert \psi _{a}^{\alpha }\right\rangle =\left\langle
\psi _{b}^{\alpha }\right\vert K_{a}^{\alpha }\left\vert \psi _{b}^{\alpha
}\right\rangle =K_{ba}^{\alpha \alpha }$, plus the analog term for $%
K_{ab}^{\beta \beta }$; observe that there is no $K_{ab}^{\alpha \beta }$
term since such interaction does not exist. Furthermore, the explicit
expression for each one of these integrals is straightly obtained from the
spatial contributions given in Eq. (\ref{integrals}). As should be expected,
Eq. (\ref{eo1}) reduces to Eq. (\ref{E0_fe}) when $\psi _{j}^{\alpha }=\psi
_{j}^{\beta }$.

The Pople-Nesbet formalism has analogy to the Roothaan one. The difference
is that, instead of Eq. (\ref{exp}), there are two distinct expansions for
the spatial orbitals $\psi ^{\alpha }$ and $\psi ^{\beta }$ in terms of the
known basis functions $\left\{ \phi _{\nu }|\nu =1...k\right\} $, given by 
\begin{equation}
\psi _{i}^{\alpha /\beta }=\sum_{\nu }C_{\nu i}^{\alpha /\beta }\phi _{\nu }%
\text{.}  \label{exsp}
\end{equation}%
By following the same procedure as in the Roothaan matrix expressions, one
obtains the two characteristic $k$ x $k$ coupled equations for the
Pople-Nesbet approach, 
\begin{equation}
\mathbf{F}^{\alpha /\beta }\mathbf{C}^{\alpha /\beta }=\mathbf{SC}^{\alpha
/\beta }\mathbf{\varepsilon }^{\alpha /\beta }\text{.}  \label{pn}
\end{equation}%
Here the overlap matrix $\mathbf{S}$ is defined in Eq. (\ref{over}), and the
meanings of $\mathbf{C}^{\alpha /\beta }$, $\mathbf{\varepsilon }^{\alpha
/\beta }$, and $\mathbf{F}^{\alpha /\beta }$ become understood from the
discussion following that equation. The respective Fock matrices are given
by 
\begin{equation}
F_{\mu \nu }^{\alpha /\beta }=\int d^{3}\mathbf{r}\phi _{\mu }^{\ast }(%
\mathbf{r})f^{\alpha /\beta }\phi _{\nu }(\mathbf{r})\text{.}  \label{irres}
\end{equation}%
It is also convenient in this UHF formalism for open shell systems to
introduce the charge density of the up and down spin electrons, $\rho
^{\alpha /\beta }(\mathbf{r})=\sum_{a}^{N^{\alpha /\beta }}\left\vert \psi
_{a}^{\alpha /\beta }(\mathbf{r})\right\vert ^{2}=\sum_{\mu }\sum_{\nu
}P_{\mu \nu }^{\alpha /\beta }\phi _{\mu }(\mathbf{r})\phi _{\nu }^{\ast }(%
\mathbf{r})$, where the elements of the corresponding density matrix for the
up and down states are 
\begin{equation}
P_{\mu \nu }^{\alpha /\beta }=\sum_{a}^{N^{\alpha /\beta }}C_{\mu a}^{\alpha
/\beta }C_{\nu a}^{\alpha /\beta \ast }\text{.}  \label{ppp}
\end{equation}%
From these expressions one can define two new quantities. One is the total
charge density, $\rho ^{T}(\mathbf{r})=\rho ^{\alpha }(\mathbf{r})+\rho
^{\beta }(\mathbf{r})$, which yields the total number of carriers, $N$,
after integrated over all space; the other is the spin density, $\rho ^{S}(%
\mathbf{r})=\rho ^{\alpha }(\mathbf{r})-\rho ^{\beta }(\mathbf{r})$, whose
integral over all space yields $2M_{S}$. The latter shows that unrestricted
wavefunctions are eigenfunctions of $S_{Z}$, but not necessarily of $S^{2}$.
Consequently, one can define the total, $\mathbf{P}^{T}$, and spin, $\mathbf{%
P}^{S}$, density matrices for the system as 
\begin{gather}
\mathbf{P}^{T}=\mathbf{P}^{\alpha }+\mathbf{P}^{\beta }\text{,}  \notag \\
\mathbf{P}^{S}=\mathbf{P}^{\alpha }-\mathbf{P}^{\beta }\text{.}  \label{popu}
\end{gather}

Furthermore, the elements of the two Fock matrices can be obtained by
inserting Eq. (\ref{fafb2}) into Eq. (\ref{irres}), and using Eqs. (\ref%
{exsp}), (\ref{ppp}) and (\ref{popu}). After some algebra, one finds $F_{\mu
\nu }^{\alpha /\beta }=T_{\mu \nu }+G_{\mu \nu }^{\alpha /\beta }$, where
the kinetic matrix $T_{\mu \nu }$ is defined in Eq. (\ref{kin}), and the two
Coulomb matrices $G_{\mu \nu }^{\alpha /\beta }$ are given by 
\begin{gather}
G_{\mu \nu }^{\alpha /\beta }=\frac{e^{2}}{\varepsilon }\sum_{\lambda
}\sum_{\sigma }\left[ P_{\lambda \sigma }^{T}\int d^{3}\mathbf{r}_{1}\int
d^{3}\mathbf{r}_{2}\phi _{\mu }^{\ast }(\mathbf{r}_{1})\phi _{\sigma }^{\ast
}(\mathbf{r}_{2})\frac{1}{\left\vert \mathbf{r}_{1}-\mathbf{r}%
_{2}\right\vert }\phi _{\nu }(\mathbf{r}_{1})\phi _{\lambda }(\mathbf{r}%
_{2})\right.   \notag \\
\left. -P_{\lambda \sigma }^{\alpha /\beta }\int d^{3}\mathbf{r}_{1}\int
d^{3}\mathbf{r}_{2}\phi _{\mu }^{\ast }(\mathbf{r}_{1})\phi _{\sigma }^{\ast
}(\mathbf{r}_{2})\frac{1}{\left\vert \mathbf{r}_{1}-\mathbf{r}%
_{2}\right\vert }\phi _{\lambda }(\mathbf{r}_{1})\phi _{\nu }(\mathbf{r}_{2})%
\right] \text{.}  \label{gab}
\end{gather}%
Notice that the self-consistency again lies in the fact that both $\mathbf{F}
$ matrices depend on the $\mathbf{P}$ matrices which, in turns, depend on $%
\mathbf{C}$ matrices. Finally, the coupling between up and down spin
equations appears on the property that $\mathbf{F}^{\alpha }$ ($\mathbf{F}%
^{\beta }$) depends on $\mathbf{P}^{\beta }$ ($\mathbf{P}^{\alpha }$)
through $\mathbf{P}^{T}$.

The procedure for the solution of Eq. (\ref{pn}) is the same used for Eq. (%
\ref{bic}), having in mind that now a simultaneous solution must be found
for the two distinct spatial orbital sets. If the criterion for convergence
in the Pople-Nesbet formalism is the ground state energy, it can be written
as 
\begin{equation}
E_{0}^{UHF}=\frac{1}{2}\sum_{\mu }\sum_{\nu }\left[ P_{\nu \mu }^{T}T_{\mu
\nu }+P_{\nu \mu }^{\alpha }F_{\mu \nu }^{\alpha }+P_{\nu \mu }^{\beta
}F_{\mu \nu }^{\beta }\right] \text{.}  \label{ea_U}
\end{equation}

As already mentioned, one disadvantage of this approach is that unrestricted
functions, in general, are not eigenstates of the total spin $S$, even
though they have defined $S_{Z}$ values. However, one may use the expressions%
\cite{szabo} 
\begin{equation}
\left\langle S^{2}\right\rangle _{UHF}=\left( \frac{N^{\alpha }-N^{\beta }}{2%
}\right) \left( \frac{N^{\alpha }-N^{\beta }}{2}+1\right) +N^{\beta
}-\sum_{a}^{N^{\alpha }}\sum_{b}^{N^{\beta }}\left[ \sum_{\mu }\sum_{\nu
}C_{\mu a}^{\alpha \text{ }\ast }C_{\nu b}^{\beta }S_{\mu \nu }\right] ^{2}%
\text{ }  \label{S2}
\end{equation}%
in order to get an estimative for $S$, while $S_{Z}$ is obtained from 
\begin{equation}
\left\langle S_{Z}\right\rangle _{UHF}=\frac{1}{2}\sum_{\mu }\sum_{\nu
}\left( P_{\nu \mu }^{\alpha }-P_{\nu \mu }^{\beta }\right) S_{\mu \nu }%
\text{.}  \label{SZ}
\end{equation}%
As a last observation, the expected value of $S$ will, in general, be higher
because of the contamination from other symmetries.

\section{Spherical quantum dot}

As an application of both RHF and UHF formalisms, we consider a QD with
radius $R_{0}$ confined by an infinite spherical potential in the presence
of a magnetic field $\mathbf{B}=B_{0}(0,0,1)$ that can be populated up to $40
$ electrons. The single-particle Hamiltonian has the form 
\begin{equation}
H_{0}=\frac{\hbar ^{2}}{2m}\left( \mathbf{k}+\frac{e}{\hbar c}\mathbf{A}%
\right) ^{2}+g\frac{\mu _{B}}{\hbar }\mathbf{B}\cdot \mathbf{S}\text{,}
\label{H1P}
\end{equation}%
where $\mu _{B}=e\hbar /(2m_{0}c)$ is the Bohr magneton, $g$ is the bulk $g$%
-factor, and the vector potential is used in the symmetric gauge $\mathbf{A}%
=(\mathbf{B}\times \mathbf{r})/2$. By using the atomic units, $%
E_{Ry}=e^{2}/(2a_{0})$ for energy and $a_{0}=\hbar ^{2}/(m_{0}e^{2})$ for
length, the Hamiltonian $H_{0}$ can be written as 
\begin{equation}
H_{0}=\frac{1}{\widetilde{m}}\frac{a_{0}^{2}}{R_{0}^{2}}\left[ -\frac{1}{%
x^{2}}\frac{\partial }{\partial x}\left( x^{2}\frac{\partial }{\partial x}%
\right) +\frac{\mathbf{L}^{2}}{x^{2}}+\frac{R_{0}^{2}}{2l_{B}^{2}}\left(
L_{Z}+\widetilde{m}gS_{Z}\right) +\frac{R_{0}^{4}}{4l_{B}^{4}}x^{2}\sin
^{2}(\theta )\right] \text{,}  \label{Ha0}
\end{equation}%
where $\widetilde{m}=m/m_{0}$, $l_{B}=\sqrt{\hbar c/(eB_{0})}$ is the
magnetic length, and $x=r/R_{0}$ is a dimensionless variable. Without
magnetic field, the normalized spatial eigenfunctions of $H_{0}$ are given
by 
\begin{equation}
\phi _{\nu }(\mathbf{r})=\phi _{n,l,m_{l}}(x,\theta ,\phi )=\left[ \frac{2}{%
R_{0}^{3}}\frac{1}{\left[ j_{l+1}\left( \alpha _{nl}\right) \right] ^{2}}%
\right] ^{1/2}j_{l}\left( \alpha _{nl}x\right) Y_{l,m_{l}}(\theta ,\phi )%
\text{.}  \label{est1p}
\end{equation}%
The boundary condition at the surface $r=R_{0}$ determines $\alpha _{nl}$ as
the $n^{th}$ zero of the spherical Bessel function $j_{l}(\alpha _{nl}x)$;
also, the spherical harmonic $Y_{l,m_{l}}(\theta ,\phi )$ is the well known
eigenstate of $\mathbf{L}^{2}$ and $L_{Z}$.

\begin{table}[tbp]
\caption{Values of the five coefficients and exponents that optimize the set
of Gaussians used in the expansions for the six orbitals considered in this
work.}
\label{tab1}\centering%
\begin{tabular}{|c||c|c|c|c|c|}
\hline
$orbital$ & $V_{1},$ $D_{1}(10^{-4})$ & $V_{2},$ $D_{2}(10^{-4})$ & $V_{3},$ 
$D_{3}(10^{-4})$ & $V_{4},$ $D_{4}(10^{-4})$ & $V_{5},$ $D_{5}(10^{-4})$ \\ 
\hline\hline
$1s$ & $0.3229,$ $0.003$ & $0.2353,$ $0.328$ & $0.4554,$ $0.043$ & $0.9069,$ 
$0.330$ & $0.8317,$ $0.006$ \\ \hline
$1p$ & $0.5902,$ $0.762$ & $0.0646,$ $0.564$ & $0.5941,$ $0.447$ & $0.4599,$ 
$0.350$ & $0.9727,$ $0.724$ \\ \hline
$1d$ & $0.7301,$ $1.120$ & $0.9570,$ $0.915$ & $0.8227,$ $1.620$ & $0.1950,$ 
$0.492$ & $0.4194,$ $1.100$ \\ \hline
$2s$ & $0.4823,$ $-2.260$ & $0.9875,$ $-2.440$ & $0.1082,$ $-1.530$ & $%
0.4006,$ $-2.840$ & $0.0342,$ $-2.260$ \\ \hline
$1f$ & $0.6377,$ $1.500$ & $0.4539,$ $1.920$ & $0.9730,$ $1.440$ & $0.4921,$ 
$1.610$ & $0.5498,$ $1.360$ \\ \hline
$2p$ & $0.8513,$ $-1.420$ & $0.0413,$ $-3.720$ & $0.0565,$ $-0.401$ & $%
0.8797,$ $-1.410$ & $0.0457,$ $-1.700$ \\ \hline
\end{tabular}%
\end{table}

The Hamiltonian for the electron-electron interaction, in units of $E_{Ry}$,
has the form 
\begin{equation}
H_{ee}=\frac{2}{\varepsilon }\frac{a_{0}}{R_{0}}\frac{1}{\left\vert \mathbf{x%
}_{1}-\mathbf{x}_{2}\right\vert }\text{.}  \label{vee}
\end{equation}%
By using the multipole expansion, 
\begin{equation}
\frac{1}{|\mathbf{x}_{1}-\mathbf{x}_{2}|}=\sum_{\kappa =0}^{\infty }\frac{%
4\pi }{2\kappa +1}\frac{x_{<}^{\kappa }}{x_{>}^{\kappa +1}}\sum_{m_{\kappa
}=-\kappa }^{\kappa }\left( -1\right) ^{m_{\kappa }}Y_{\kappa ,m_{\kappa
}}\left( \theta _{1},\phi _{1}\right) Y_{\kappa ,-m_{\kappa }}\left( \theta
_{2},\phi _{2}\right) \text{,}  \label{multi}
\end{equation}%
all the angular part of the problem can be solved analytically and, after
inserted into our numerical code, we are left to solve only the radial
degree of freedom.

Without magnetic field, we take into account in Eqs. (\ref{exp}) and (\ref%
{exsp}) the spatial orbitals that define the six lowest energy shells ($1s$, 
$1p$, $1d$, $2s$, $1f$, $2p$) under this symmetry.\cite{walecka} Therefore,
the index $\nu \equiv n,l,m_{l}$ can assume up to $40$ ($20$ spin-up and $20$
spin-down) possible values for the states within those shells. Certainly,
the inclusion of a magnetic field lifts both spin and orbital degeneracies
of those states. In our application, we shall consider two possible
materials forming the QD, namely GaAs (wide-gap material) and InSb
(narrow-gap material), whose defining parameters are $\widetilde{m}=0.065$, $%
g=0.45$ and $\varepsilon =12.65$ for GaAs, and $\widetilde{m}=0.013$, $%
g=-53.1$ and $\varepsilon =16.5$ for InSb.

Due to the presence of a magnetic field, some modifications on both RHF and
UHF expressions must be made. In the Roothaan approach, the quadratic term
in $B_{0}$ ($\sim l_{B}^{-4}$) must be added to the single-particle
contribution present in $h_{aa}$ in Eq. (\ref{integrals}), and to the matrix 
$T_{\mu \nu }$ in Eq. (\ref{kin}); the linear terms in $B_{0}$ ($\sim
l_{B}^{-2}$) are zero for a closed shell configuration, thus yielding no
contribution. However, all terms proportional to $B_{0}$ have to be
considered in the Pople-Nesbet formalism, since we will be dealing with an
open shell configuration. Both linear ($\sim B_{0}L_{Z}$) and quadratic
(diamagnetic) terms can be straightly added to the definitions of $h_{aa}$
(Eq. (\ref{integrals})) and $T_{\mu \nu }$ (Eq. (\ref{kin})). Moreover, the
inclusion of the spin-dependent linear term ($\sim B_{0}S_{Z}$) in $h_{aa}$
will impose the kinetic matrix $T_{\mu \nu }$ be decomposed in its
components $T_{\mu \nu }^{\alpha /\beta }$, as it occurs with $G_{\mu \nu
}^{\alpha /\beta }$ in Eq. (\ref{gab}). Thus, under a magnetic field, we
must make in the ground state energy of an unrestricted system, given in Eq.
(\ref{ea_U}), the substitution $P_{\nu \mu }^{T}T_{\mu \nu }=P_{\nu \mu
}^{\alpha }T_{\mu \nu }^{\alpha }+P_{\nu \mu }^{\beta }T_{\mu \nu }^{\beta }$%
.

\begin{figure}[tbp]
\includegraphics[width=1.0\linewidth]{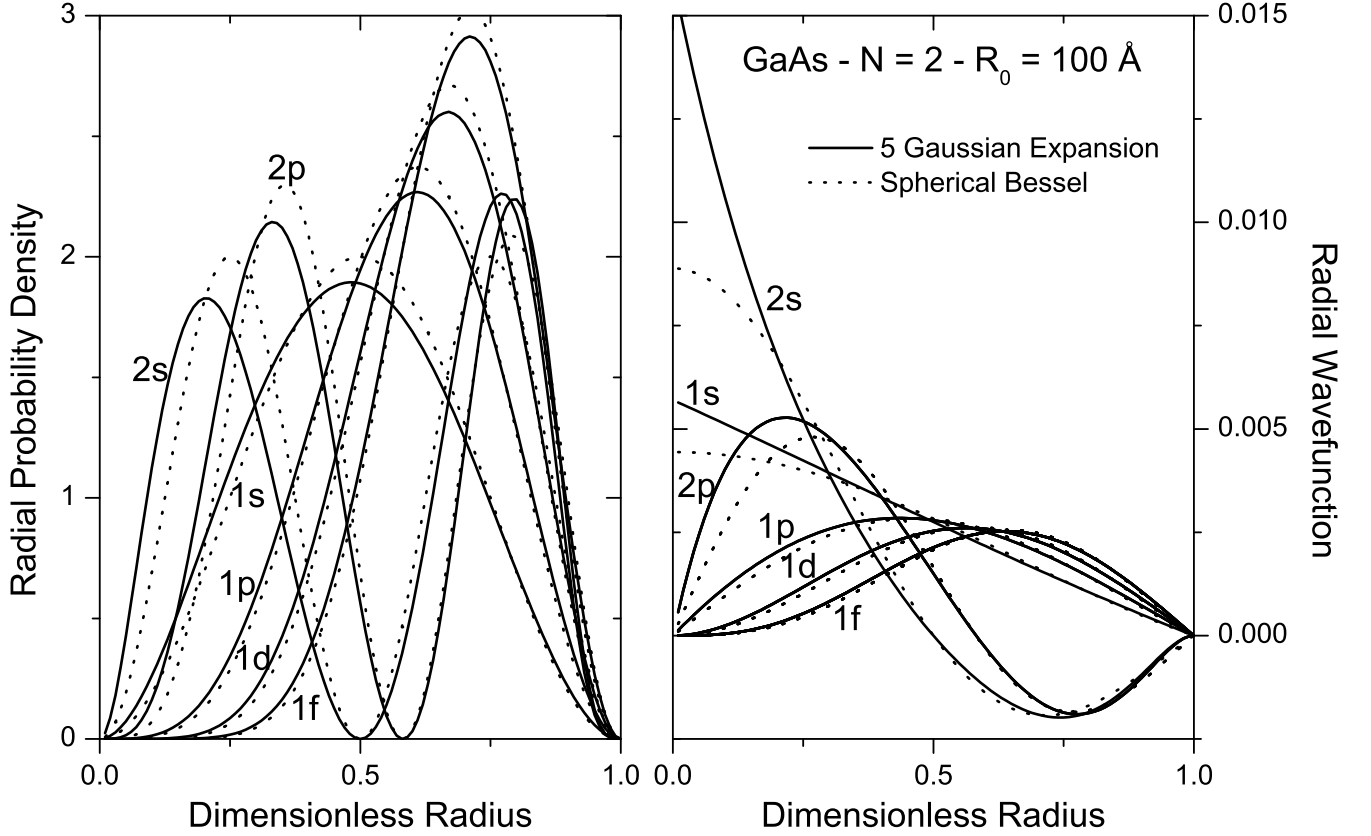} 
\caption{Comparison between the expanded Gaussian basis sets used in our
calculations and the respective exact spherical Bessel functions, for a GaAs
QD having $N=2$ and $R_{0}=100$ \AA . The left panel shows the probability
densities, while the right one shows the wavefunctions. The agreement is
good for all orbitals except very closed to the origin for $s$-states.}
\label{fig1}
\end{figure}

The last important detail in our approach refers to the orbital basis $%
\left\{ \phi _{\nu }|\nu =1...k\right\} $ used in our calculations. Instead
of the spherical Bessel functions of Eq. (\ref{est1p}), the radial part of
each orbital is decomposed in a sum involving five Gaussians confined to a
sphere of radius $R_{0}$, while the angular part is maintained as defined by
its symmetry. So, we change the basis in Eq. (\ref{est1p}) by 
\begin{equation}
\phi _{n,l,m_{l}}(x,\theta ,\phi )=N_{nl}\left( 1-x\right)
^{n}x^{l}\prod_{i=1}^{n-1}\left( \widetilde{\alpha }_{il}-x\right)
\sum_{k=1}^{5}V_{k}e^{-D_{k}R_{0}^{2}x^{2}}Y_{l,m_{l}}(\theta ,\phi )\text{,}
\label{gaus}
\end{equation}%
where $N_{nl}$ is the orbital normalization, the polynomial in $\left(
1-x\right) $ satisfies the boundary condition at $x=1$ ($r=R_{0}$), and the
polynomial in $x$ makes the functions having $l>0$ be zero at the origin $x=0
$; also, the product in $\left( \widetilde{\alpha }_{il}-x\right) $ makes
the function be zero at the zeros $\widetilde{\alpha }_{il}$ of the
respective spherical Bessel function transposed to the interval $0\leq x\leq
1$, and the last sum involves an expansion in five Gaussians. For $N\leq 40$%
, higher number of Gaussians in the expansion does not show any improvement
on our results. The coefficients $V_{k}$ as well as the exponents $D_{k}$
are determined for each value of $R_{0}$, and are obtained by using the
condition of maximizing the superposition between Eq. (\ref{gaus}) and the
respective spherical Bessel function. Once $V_{k}$ and $D_{k}$ are
determined and the basis is found, we run our RHF (UHF) code for given
values of $R_{0}$ and $N$, and find the parameters $C_{\nu i}$ ($C_{\nu
i}^{\alpha /\beta }$) that better describe Eq. (\ref{exp}) (Eq. (\ref{exsp}%
)) and give the minimal energy in Eq. (\ref{E0_root2}) (Eq. (\ref{ea_U})).

Finally, we would like to calculate two closely related quantities that give
interesting information on the charging processes of a confined system. The
first one is the QD chemical potential, $\mu _{dot}(N)$, which yields the
energy difference between two successive ground states, and can be
calculated as 
\begin{equation}
\mu _{dot}(N)=E_{0}(N)-E_{0}(N-1)\text{.}  \label{pqui}
\end{equation}%
The second one is the QD charging energy, $E_{char}(N)$, which yields the
energy cost for the addition of an extra electron to the system, 
\begin{equation}
E_{char}(N)=I(N)-A(N)=E_{0}(N+1)-2E_{0}(N)+E_{0}(N-1)\text{.}  \label{ecar}
\end{equation}%
Here, $I(N)=E_{0}(N-1)-E_{0}(N)$ is the ionization potential, while $%
A(N)=E_{0}(N)-E_{0}(N+1)$ is the electronic affinity. It becomes clear, from
these two equations, that $E_{char}(N)=\mu _{dot}(N+1)-\mu _{dot}(N)$.

\section{Results}

Figure \ref{fig1} shows a comparison between the exact orbitals described by
the spherical Bessel functions of Eq. (\ref{est1p}) and the expansions
involving the Gaussians of Eq. (\ref{gaus}) used in our calculations for a
GaAs QD having $N=2$ and $R_{0}=100$ \AA . The right panel shows the radial
wavefunctions while the left one shows the radial probability densities.
Although one can observe a difference in the $s$-wavefunctions, near $x=0$,
their probability densities are reasonable in that region. For any other
occupation and radius, as also for InSb QDs, this same feature regarding $s$%
-orbitals is observed. Table \ref{tab1} shows the optimized coefficients and
exponents for the five Gaussians related to the six orbitals taken into
account. The factor $10^{-4}$ in all exponents cancels the term $R_{0}^{2}$
in Eq. (\ref{gaus}). Besides, all exponents related to any orbital having $%
n=2$ are negative since there are regions where those wavefunctions possess
negative values.

\begin{figure}[tbp]
\includegraphics[width=1.0\linewidth]{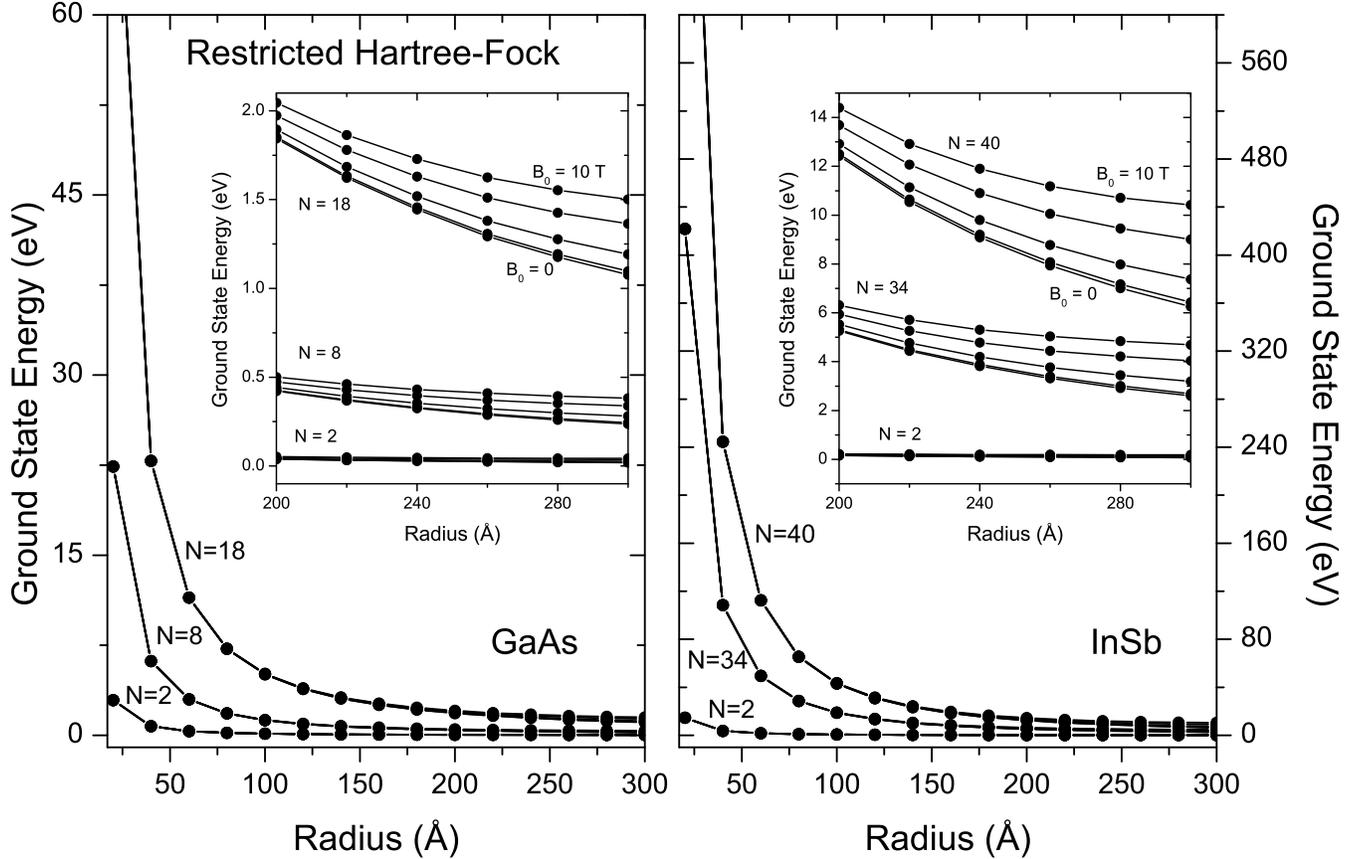} 
\caption{Restricted ground state energies for a GaAs (left panel) and InSb
(right panel) QD populated with $N=2$, $8$ and $18$ ($N=2$, $34$ and $40$)
electrons. The Coulomb contribution is more important at larger radii. In
the insets we analyze the influence of a magnetic field in the system, where
one can see that even in a closed shell configuration having $L=S=0$, the
presence of the field is visible due to the diamagnetic term.}
\label{fig2}
\end{figure}

In figure \ref{fig2} we show results of a RHF-Roothaan calculation. In the
left panel we present the ground state energy as function of the radius for
a GaAs QD populated with $2$, $8$ and $18$ electrons, while the right one
shows the same, but for a InSb QD populated with $2$, $34$ and $40$
electrons. These five distinct values, plus $N=20$, are the successive magic
numbers for the closed shell configurations at this symmetry. In both QDs,
it is noticed that the kinetic energy is much higher and totally dominates
the Coulomb one at smaller radii. Thus, the energies related to different $N$
values are more distant from each other. At larger radii, or smaller
electronic density, the electron-electron interaction becomes more important
and the energy separation decreases. In the insets, which show a zoom at
larger radii, we analyze the influence of the magnetic field on the system
shell configuration. For both occupations $N$, the curves for increasing
energy ordering refer to fields of $0$, $2$, $5$, $8$ and $10$ T, even
though we have only labeled the cases with $N=18$ and $N=40$. For $N=2$ the
presence of the magnetic field is imperceptible at those inset scales. As
also expected, due to its large $g$-factor, the Zeeman splitting is much
higher for the InSb QD (observe the different energy scales in the insets).
Most interesting is the fact that even for a closed shell configuration,
where $L=S=0$, the influence of the quadratic field (diamagnetic term) of
Eq. (\ref{Ha0}, which is the only non-zero contribution in the restricted
case, becomes important mainly for higher occupation numbers and fields, as
well as at larger radii (notice that there are almost no difference between $%
0$ and $2$ T for any $N$).

\begin{figure}[tbp]
\includegraphics[width=0.90\linewidth]{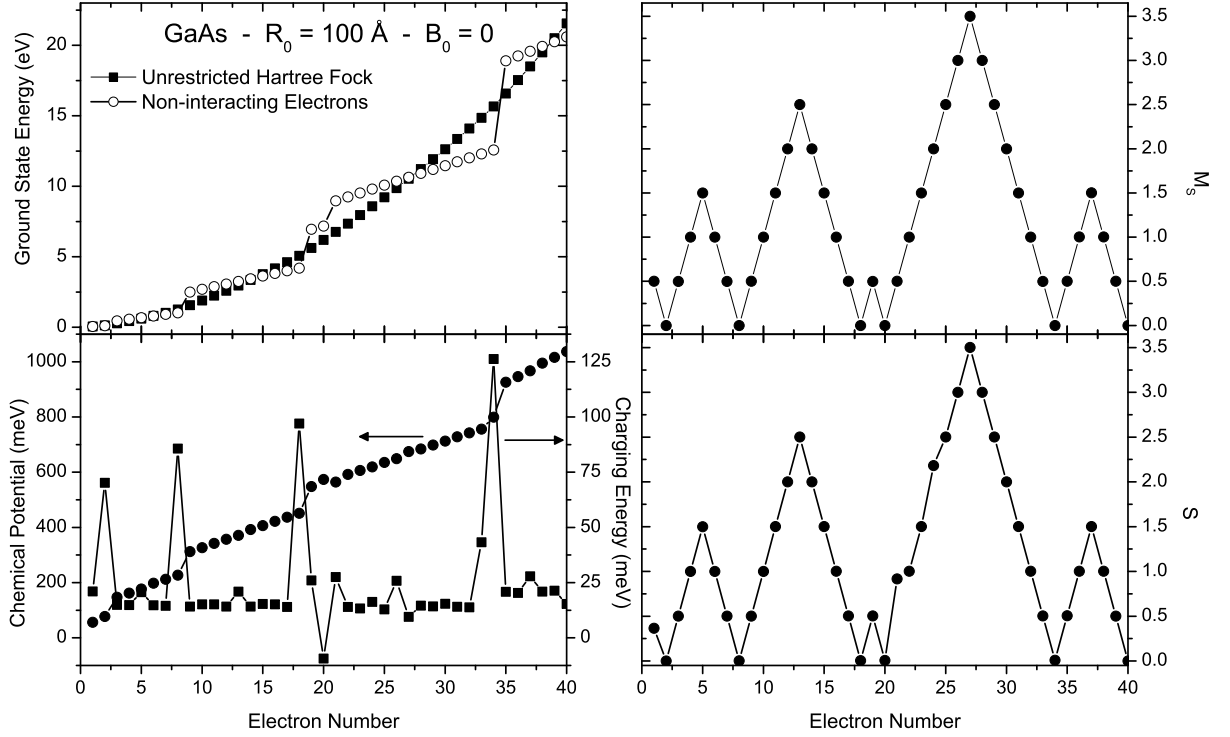} 
\caption{Unrestricted ground state energies for a $R_{0}=100$ \AA\ GaAs QD
without magnetic field. In the upper left panel we compare the unrestricted
and the non-interacting energies, where the QD energy shell structure is
visible. The bottom left panel shows QD chemical potential (left scale) and
charging energy (right scale); the former displays abrupt change always that
a new shell starts to be populated, while the latter presents larger
(smaller) peaks when a shell is filled (half-filled), a direct consequence
of the Hund rule. The bottom and upper right panels show, respectively, the $%
N$-evolution of the expected values of total spin $S$ and its projection $%
M_{S}$.}
\label{fig3}
\end{figure}

In order to study spectra for any occupation we show, in figure \ref{fig3},
the results of a UHF-Pople-Nesbet calculation for a GaAs QD having $R_{0}=100
$ \AA\ and without magnetic field. In the left upper panel it is made a
comparison between the UHF results and the non-interacting electron case,
where the energy shell structure is clear for $N=2$, $8$, $18$, $20$, $34$
and $40$. Notice that the electron-electron interaction makes the energy of
a QD, whose occupation corresponds to a shell less (more) than half-filled,
be decreased (increased) with respect to the non-interacting case. When such
occupation corresponds exactly to half-filled cases ($N=5$, $13$, $27$ and $%
37$), the interacting energy is approximately equal to the non-interacting
one.

In the left bottom panel of figure \ref{fig3} we show both QD chemical
potential (left scale, Eq. (\ref{pqui})) and charging energy (right scale,
Eq. (\ref{ecar})), where the respective values of $E_{0}$ are obtained from
the unrestricted calculation presented in the left upper panel. Notice that $%
\mu _{dot}$ increases linearly as the occupation gets higher inside a given
shell. When such shell is totally filled, there is an abrupt change in $\mu
_{dot}$ indicating that the following shell starts its occupation; observe
that the higher the occupation, the larger is the changing. An anomalous
behavior seems to occur for the $2s$ shell, whose $\mu _{dot}$ value is
larger than the one for the $1f$ shell (that has higher energy). The
charging energy is another way to verify not only the presence of shell
structure for the spectrum, but also the validity of Hund's rule for the
filling of such shells. In principle, $E_{char}$ must present larger
(smaller) peaks when the total (half) occupation of a given shell is
achieved; the first fact is due to the higher difficulty to the addition of
an extra electron to a QD when a filled shell state is reached. The second
one refers to Hund's rule, which establishes that electrons must be added to
the system with their spins being parallel, until all possible orbitals
inside a given shell be occupied; this makes the total energy of the system
smaller since this procedure maximizes the negative exchange contribution.
However, some violations of this rule can be verified in $E_{char}$: the
smaller peak of $N=27$ occurs here at $N=26$, and the larger peak of $N=20$
has a negative value.

The bottom and upper right panels of figure \ref{fig3} show respectively the
evolution of the total spin $S$ and its projection $M_{S}$ as a function of
QD population, calculated from Eqs. (\ref{S2}) and (\ref{SZ}) for the
unrestricted energies. Notice that, without magnetic field, the Hund rule
seems to be followed for all states with up to 40 electrons. The $M_{S}$
expected value oscillates from $0$ in a filled shell to its maximum in a
half-filled shell, when it starts to decrease again on the way to the
closing of the shell; the maxima are $M_{S}=1/2$, $3/2$, $5/2$ and $7/2$ for 
$s$, $p$, $d$ and $f$ shells, respectively. The $S$ expected value yielded
by the unrestricted formalism is also very reasonable; discrepancies are
only observed at $N=24$, where $S>2$, and at $N=21$, where $S>1/2$. We
believe that both discrepancies related to the $2s$ shell or to its
surroundings - $\mu _{dot}$ larger than the one of $1f$ shell, negative peak
for $N=20$ in $E_{char}$, and almost doubled $S$ expected value for $N=21$ -
are caused by the non-reasonable Gaussian reproduction of this orbital, as
visible in figure \ref{fig1}. These same qualitative results are observed
for an InSb QD without magnetic field.
\begin{figure}[tbp]
\includegraphics[width=0.75\linewidth]{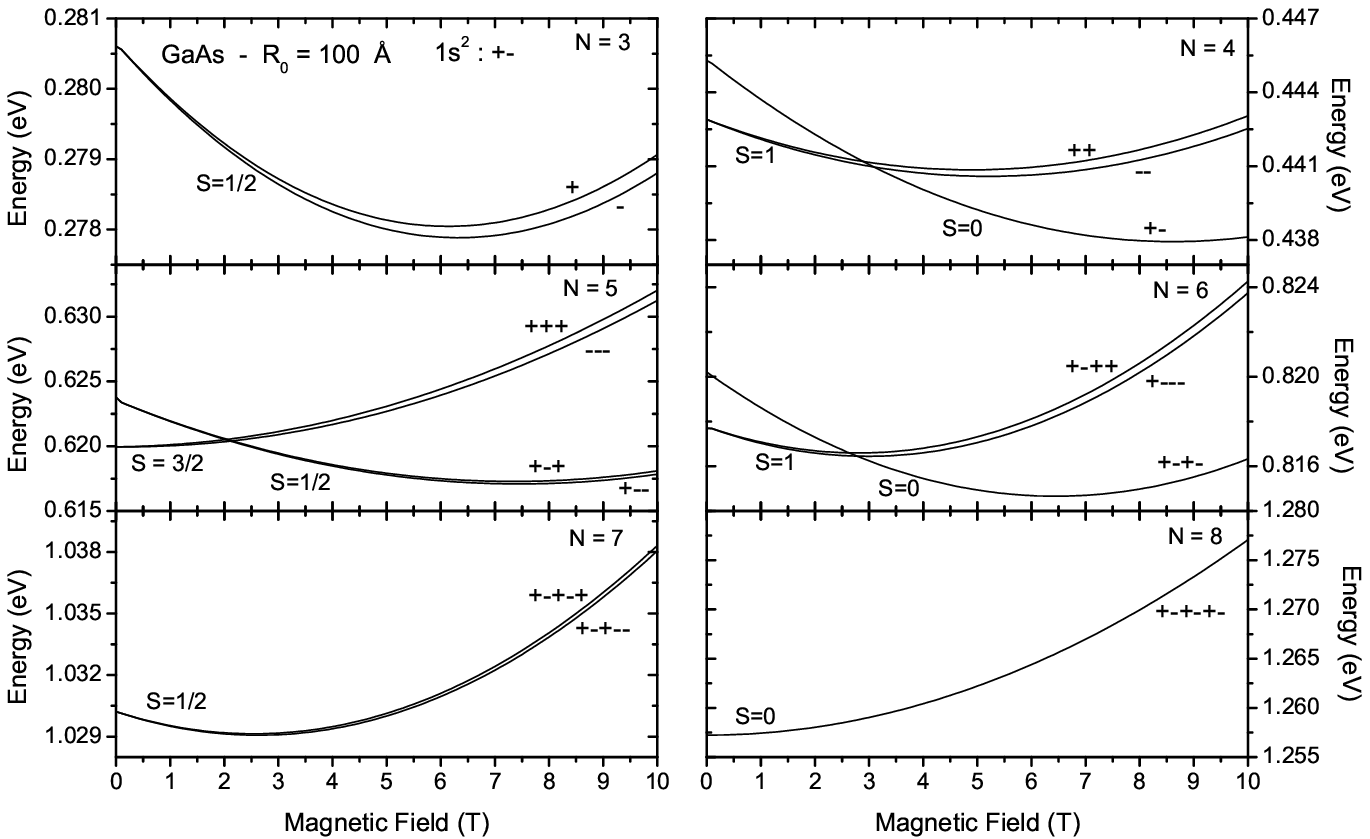} 
\caption{Violation of Hund's rule induced by magnetic field in the $R_{0}=100
$ \AA\ GaAs QD of the previous figure. The panels show the successive
occupation (indicated in the upper right corner of each panel) of the $1p$
shell, assuming that the $1s$ shell remains populated by one spin-up and one
spin-down electron. The possible spin configurations for given $N$ are
indicated by $+$ (spin-up) and $-$ (spin-down). For $B_{0}=0$ the spin
sequence is $1/2-1-3/2-1-1/2-0$, while at fields higher than $3$ T it
changes to $1/2-0-1/2-0-1/2-0$. It is interesting to observe that a magnetic
field is able to suppress the energy triplets and quartets of $p$-shells
from the QD spectrum.}
\label{fig4}
\end{figure}

By focusing on the $1p$ shell we show in figure \ref{fig4}, for the same QD
of the previous figure, how a finite magnetic field is able to violate
Hund's rule in the system. Panels from left to right and from up to bottom
show the successive ground state energies from $N=3$ to $N=8$ as this shell
is filled, always considering that the $1s$ shell remain fully occupied by
two electrons, one spin-up and one spin-down; the distinct possible spin
configurations for each $N$ are indicated by $+$ (spin-up) and $-$
(spin-down). In addition to the small Zeeman effect present in all
occupations, there is a change of ground state spins at $N=4$, $5$ and $6$
as the field is increased. Notice that at zero field the spin sequence is $%
1/2-1-3/2-1-1/2-0$; in a field above $3$ T it becomes $1/2-0-1/2-0-1/2-0$,
meaning that quartets and triplets are suppressed by the magnetic field, and
the ground state of the system starts to oscillate only between singlets and
doublets at high fields as $N$ increases. When this $1p$ shell is
half-filled ($N=5$), the ground state goes from a quartet to a doublet at $%
B_{0}\simeq 2$ T; when it has one electron more ($N=6$) or less ($N=4$) than
that, it goes from a triplet to a singlet at $B_{0}\simeq 3$ T.

At last, we have compared the results from both RHF and UHF self-consistent
matrix formulations with the ones obtained from the $\mathbf{LS}$-coupling
scheme used in Ref. [\onlinecite{LS}], where a GaAs QD having $R_{0}=90$ 
\AA\ was considered, and the quadratic term in $B_{0}$ was neglected since
only small magnetic fields were considered. Also, only $N=2$ and $N=3$
occupations were calculated, since the states were constructed analytically
(not only a single Slater determinant), and the electron-electron
interaction was included by using perturbation theory, that is justified at
such radius. At zero field the energies for $N=2$ are $16.5$ meV ($\mathbf{LS%
}$) and $16.1$ meV (RHF), while for $N=3$ they are $34.8$ meV ($\mathbf{LS}$%
) and $33.9$ meV (UHF). Therefore, both Roothaan and Pople-Nesbet formalisms
indeed give smaller ground state energies than the $\mathbf{LS}$
perturbative scheme. We have also checked the validity of neglecting the
diamagnetic term (quadratic in $B_{0}$) for fields smaller than $2$ T. We
may emphasize here that a disadvantage of the UHF approach is, in principle,
that one is never sure to get trustable information about the expected
values for $L$ and $S$ in a given QD state. On the other hand, the
applicability of the $\mathbf{LS}$ scheme is cumbersome and becomes very
complicated to be handled analytically as the QD occupation increases.

\section{Conclusions}

We have shown how the mean-field Roothaan and Pople-Nesbet formalisms
applied to a spherical quantum dot confined system under applied magnetic
field yield a fairly good description of its energy shell structure. For a
maximum population of $40$ electrons considered, the appropriated Gaussian
basis set for each radius has been found. We have seen how a magnetic field
influences the total energy of ground states even in closed shell
configurations. We have also shown how both chemical potential and charging
energy reproduce the closing and half-closing structures of the quantum dot
energy shells. With the calculation of the total spin expected value for
each occupation, in a given radius, we have observed that Hund's rule is
satisfied at zero field. However, under finite magnetic field, we have shown
that its applicability is violated and, at given values of the field, which
depend on quantum dot radius and material parameters, there are transitions
that change a given ground state symmetry.

\begin{acknowledgments}
This work has been supported by Funda\c{c}\~{a}o de Amparo \`{a} Pesquisa do
Estado de S\~{a}o Paulo (FAPESP) and by Conselho Nacional de Desenvolvimento
Cient\'{\i}fico e Tecnol\'{o}gico (CNPq), Brazil. Also we are in debt with
C. Trallero-Giner for long discussions on this subject.
\end{acknowledgments}

\end{document}